\documentclass[graybox]{SNmult}

\usepackage{type1cm}
\usepackage{makeidx}
\usepackage{graphicx}
\usepackage{multicol}
\usepackage[bottom]{footmisc}
\usepackage{amssymb}	
\usepackage{newtxtext}
\usepackage[varvw]{newtxmath}
\usepackage[numbers]{natbib}
\usepackage[hidelinks]{hyperref}
\usepackage{siunitx}
\usepackage{amsmath}
\usepackage{soul}
\usepackage{microtype}

\makeindex

\begin{document}
	
	\title*{Direct Numerical Simulation of Vertical-Axis Wind Turbine Near-Wake Dynamics}
	
	\author{Harry Dunn and Mohsen Lahooti}
	
	\institute{Harry Dunn \at School of Engineering, Newcastle University, Newcastle upon Tyne, UK \email{h.dunn1@newcastle.ac.uk}
		\and Mohsen Lahooti \at School of Engineering, Newcastle University, Newcastle upon Tyne, UK \email{Mohsen.Lahooti@newcastle.ac.uk}}
	
	\maketitle
	
	\abstract{Geometrically-resolved Direct Numerical Simulations of vertical-axis wind turbines are presented. Simulations were performed using the spectral/hp element method framework Nektar\texttt{++} with a moving reference frame formulation. Three turbine geometries are considered with one, two and three blades. The full dynamic stall process is resolved, including the formation of large-scale vortices, the separation from the blade, and interaction with the near wake. Increasing blade number introduces blade-vortex interactions that interact with the dynamic stall process. For the three-bladed configuration, these interactions coincide with the early phase of dynamic stall during which the laminar separation bubble develops, and the resulting dynamic stall vortex is reduced in size. Further, the DNS outcomes identified a novel complementary mechanism, where direct vortex impingement causes the premature breakup of the dynamic stall vortices. With the dynamic stall vortex a defining feature of the VAWT near-wake, its accelerated breakdown removes the flow structures that distinguish the VAWT wake from that of a bluff body. Hence the near-wake transitions more rapidly towards bluff-body dynamics, with shear-layer-associated recovery. Self-similarity analysis is extended into the near-wake to quantify this transition, capturing the downstream rate at which the wake loses its dependence on blade-generated coherent structures and collapses onto a self-similar solution. Blade number is shown to be more influential than tip-speed ratio in setting the rate of this transition. The results have implications for closely-spaced turbine arrays and coupled-pair configurations, where the inflow experienced by a downstream rotor is shown to be blade-number-dependent.}
	
	\section{Introduction}
	\label{sec:intro}

	 While most wind energy research has focused on horizontal-axis wind turbines (HAWT) due to their industrial relevance, vertical-axis wind turbines (VAWT) have attracted increasing attention due to potential benefits within offshore, urban and remote environments, as well as claims of improved scalability for major energy production \citep{dabiriPotentialOrderofmagnitudeEnhancement2011, handAerodynamicDesignPerformance2021}. Their lower centre of gravity simplifies design and maintenance, ideal in harsh and offshore floating environments. Also, their omnidirectional insensitivity to incoming flow removes the need for yaw control. The scalability derives from wake recovery characteristics, with VAWTs exhibiting shorter wake regions than their HAWT counterparts \citep{kinzelEnergyExchangeArray2012}. Combining this enhanced wake recovery with beneficial aerodynamic interference between closely-spaced counter-rotating rotors, \citet{dabiriPotentialOrderofmagnitudeEnhancement2011} suggested that VAWT wind farms may achieve power densities an order of magnitude higher than those of HAWT arrays. However, in closely-spaced turbine arrays, each rotor operates within the near-wake of nearby turbines. This near-wake is dominated by flow features associated with dynamic stall, a process inherent to VAWT operation \citep{bachantCharacterisingNearwakeCrossflow2015, brochierWaterChannelExperiments1986, posaWakeStructureSingle2016}. Dynamic stall initiates when the blades encounter rapid changes in angle of attack (AoA) which leads to leading-edge separation, followed by the roll-up of the boundary layer into a large-scale dynamic stall vortex (DSV), causing massive separation across the chord length, $c$ \citep{mccroskeyDynamicStallExperiments1976, visbalAnalysisDynamicStall2018}. The downstream DSV shedding leads to a near-wake characterised by strong interplay with these large-scale coherent structures \citep{arayaTransitionBluffbodyDynamics2017, bachantCharacterisingNearwakeCrossflow2015}. Therefore, the near-wake is strongly dependent on geometric and operational parameters influencing dynamic stall. This is a stark contrast to the bluff-body wake dynamics experienced in the far-wake \citep{arayaTransitionBluffbodyDynamics2017}. The parameters influencing dynamic stall are the tip-speed ratio, $\lambda$, (TSR, the ratio of the aerofoil's tangential velocity to the inflow freestream velocity) and the turbine static solidity $\sigma$. As the variation in AoA is TSR dependent, this parameter underlies the extent of the dynamic stall process \citep{lefouestDynamicStallDilemma2022} and drives differential blade forcing seen across the leeward and windward passages (Fig.~\ref{fig:schematic}(a)), which is strongly connected to the widely reported near-wake asymmetry \citep{battistiExperimentalBenchmarkData2018, brochierWaterChannelExperiments1986, stromNearwakeDynamicsVerticalaxis2022}. As for the solidity, increasing this parameter reduces the spacing between blades, prolonging each blade's exposure to the upstream blade's wake, thereby increasing the extent of blade-vortex interactions (BVI). Previous work on isolated aerofoils has identified BVI as interacting with dynamic stall mechanisms, specifically laminar separation bubble instability, changes in effective AoA, and boundary layer thickening \citep{barnesCounterclockwiseVorticalGustAirfoil2018, colliWindTunnelExperiments2024, zanottiStereoParticleImage2014}. This is consistent with experimental VAWT observations that the strength of the DSV shed into the near-wake decreases with increasing solidity \citep{arayaTransitionBluffbodyDynamics2017}. Capturing these dynamics therefore requires a method that resolves both the blade boundary layer and the downstream near-wake. Experimental approaches have been of great benefit in providing ground truths about these complex VAWT flows \citep{battistiExperimentalBenchmarkData2018, brochierWaterChannelExperiments1986}. However, they can be limited at capturing near-geometry and three-dimensional data \citep{simaoferreiraVisualizationPIVDynamic2009, weiNearwakeStructureFullscale2021}, as well deficiencies in replicating atmospheric conditions \citep{veersGrandChallengesScience2019}. High-fidelity large-eddy simulation (LES) has been used, but is sensitive to the subgrid model employed \citep{shamsoddinLargeEddySimulation2014}, and is often coupled with the actuator line method (ALM) \citep{abkarSelfsimilarityFlowCharacteristics2017}. The ALM does not geometrically resolve the blade and therefore cannot capture the boundary-layer development, laminar separation bubble dynamics, or dynamic stall vortex formation that dominates the rotor region and governs the near-wake. It also neglects the blade-vortex interactions which modulate these flow features for VAWT configurations.
	 
	 Here, geometrically-resolved Direct Numerical Simulations (DNS) of VAWT flow dynamics are presented. Unlike LES-ALM, this method captures the roll-up of the boundary layer into the DSV, its separation from the blade, and its convection into the near-wake. Solidity is varied across a broad range of tip-speed ratios. A novel finding of the present work is the identification of a vortex-impingement mechanism through which BVI interferes with the dynamic stall cycle and drives premature DSV breakdown. Downstream, the recovery of VAWT wakes has been characterised through self-similarity analysis, whereby the scaled velocity deficit profiles are tested for collapse onto a functional form \citep{tennekesFirstCourseTurbulence1972}. For VAWTs, this was demonstrated by \citet{abkarSelfsimilarityFlowCharacteristics2017}, though restricted to a single three-bladed configuration at two tip-speed ratios and, due to the ALM employed, to the far-wake only. Here, the analysis is performed across the parameter space and extended from the near-wake onwards, quantifying the rate at which the wake loses its dependence on blade-generated DSVs and transitions onto the self-similar solution. To the best of the authors' knowledge, this constitutes the first assessment of blade-number effects on self-similarity onset. The methodology and computational setup are detailed in Section~\ref{sec:methodology}, with results presented in Section~\ref{sec:results}.
	
	\section{Methodology}
	\label{sec:methodology}
	
		\subsection{Problem Description}
	\label{sec: problem description}
\begin{figure}[!hbtp]
		\centerline{\includegraphics[width=0.9\textwidth]{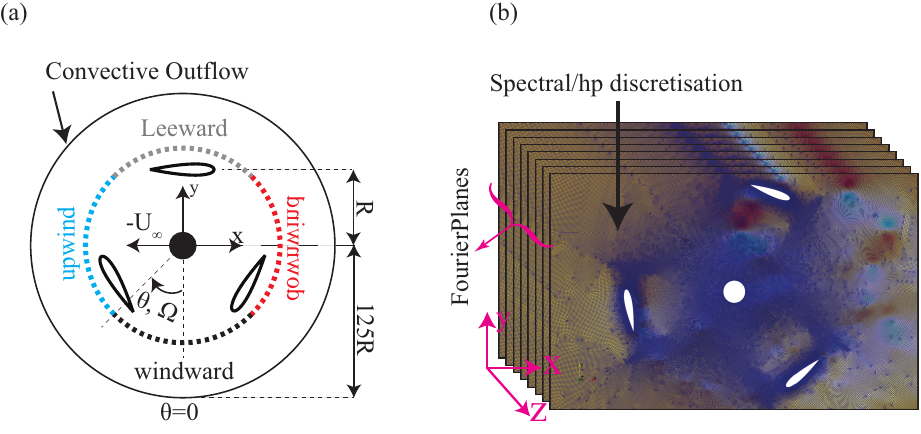}}
		\caption{VAWT Schematic: (a) top view. (b) Quasi-3D settings showing the discretisation of the domain in the $x-y$ plane using spectral/hp method and homogenous extrusion using Fourier planes in the spanwise ($z$) direction over extrusion length $L_z$.}	
		\label{fig:schematic}
	\end{figure}

	One-, two-, and three-bladed, straight-bladed Darrieus VAWT with NACA 0018 aerofoils are considered. The VAWT has a radius  $R = \SI{0.06}{m}$ and uses aerofoils with chord length $c = \SI{0.024}{m}$. The VAWT height is homogeneously extruded as $L_z = 0.2c$, with this extrusion depicted in Fig.~\ref{fig:schematic}(b). This leads to static solidities of $\sigma = \frac{N_{\text{blades}}c}{2R} = 0.2, 0.4, 0.6$ for blade numbers $N_{\text{blades}} = 1, 2, 3$ respectively. The central shaft has a radius $R_s = \SI{5e-3}{m}$ but the rotor arms are excluded due to their insignificant effects \citep{posaLargeEddySimulation2018}. VAWT operation can be characterised by the TSR, $\lambda = \frac{\Omega R}{U_\infty}$, where $\Omega$ is the VAWT's angular velocity, and $U_\infty$ is the freestream fluid velocity. Under an inflow of $U_\infty = \SI{0.15}{m/s}$ in the streamwise direction, TSR is prescribed as $\lambda = 1.5, 2.5, 4.5$ covering the range $1.5 \leq \lambda \leq 4.5$. This leads to a significant range in angle of attack (AoA), ranging as $-41.8^\circ \leq \alpha = \tan^{-1}\left(\frac{\sin \theta}{\lambda + \cos \theta}\right) \leq 41.8^\circ $. By considering both static solidity, $\sigma$, and rotation rate, a dynamic solidity can be introduced as in Eq.~\ref{eq:dynamic-solidity} \citep{arayaTransitionBluffbodyDynamics2017}.
	
	\begin{equation} \label{eq:dynamic-solidity}
		\sigma_D = 1 - \frac{R}{c}\frac{t_V}{t_{\text{conv}}} = 1 - \frac{1}{2\sigma\lambda}.
	\end{equation}
	
	In Eq.~\ref{eq:dynamic-solidity}, a characteristic length scale is defined as the sum of the gaps in the rotor circumference, $l = 2\pi R(1-\sigma)$, from which two time scales follow: (i) a VAWT time scale $t_V = l/(N_{\text{blades}} U_\infty \lambda)$, representing the time for the blades to sweep through these gaps, and (ii) a convective time scale $t_{\text{conv}} = l/U_\infty$, representing the time for the freestream to traverse the same distance. Dynamic solidity here covers the range $0.47 \leq \sigma_D \leq 0.94$, where the upper bound represents an almost solid cylinder, since from Eq.~\ref{eq:dynamic-solidity}, $\sigma_D \to 1$ as $\sigma, \lambda \to \infty$. The flow can also be characterised by a chord-based $Re_c=U_{\text{rel}}c/\nu$ where $U_{\text{rel}}=\sqrt{1+2\lambda\cos\theta+\lambda^2}$, $\nu$ is the kinematic viscosity and $\theta$ corresponds to the azimuth angle (Fig.~\ref{fig:schematic})(a)). The mean $Re_c$ over each rotation was kept to $Re_c=10^4$. The choice of $Re_c$ is similar to that seen within VAWT fundamental physics studies \citep{brochierWaterChannelExperiments1986}. Regarding applications, $Re_c \sim \mathcal{O}(10^4)$ corresponds to VAWT within urban/ extraterrestrial environments \citep{danaoExperimentalInvestigationInfluence2013, kumarLowReynoldsNumber2010}. Additionally, low $Re$ research has been used for validation within high $Re$ investigations \citep{abkarSelfsimilarityFlowCharacteristics2017, shamsoddinLargeEddySimulation2014}. Such a $Re_c$ is also critical to ensure the simulations are feasible in the context of the high-fidelity nature of the numerical setup (Section \ref{sec: numerical setup}).
	
	\subsection{Numerical Setup}
	\label{sec: numerical setup}
	DNS has been conducted for the described cases. All simulations used the high-order, spectral/hp element method discretisation within the Nektar\texttt{++} framework \citep{moxeyNektarEnhancingCapability2020}. To robustly handle the dynamics of the turbine geometry and avoid complexities associated with dynamic remeshing for handling moving geometries, the governing Incompressible Navier-Stokes and continuity equations are expressed within a Moving Reference Frame (MRF) formulation (Eq.~\ref{eq:MRF}) \citep{dunnVortexDynamicsVerticalAxis2024, lahootiDNSFluidstructureInteraction2023}.
	
		\begin{subequations}
		\label{eq:MRF}
		\begin{align}
			\frac{\partial \mathbf{u}}{\partial t} + (\mathbf{u} - \mathbf{u}_0 - \boldsymbol{\Omega}_0 \times \mathbf{x}) \cdot \nabla \mathbf{u} + \boldsymbol{\Omega}_0 \times \mathbf{u} &= -\frac{1}{\rho}\nabla p + \nu \nabla^2 \mathbf{u} + \mathbf{f} \label{eq:MRF-a} \\
			\nabla \cdot \mathbf{u} &= 0 \label{eq:MRF-b}
		\end{align}
	\end{subequations}
	
	In Eq.~\ref{eq:MRF}, $\mathbf{u}=(u,v,w)$ and $p$ are the absolute flow velocities and pressure in the MRF respectively, $\mathbf{x}=(x,y,z)$ is the position vector in the MRF, $\mathbf{u}_0=(-U_\infty,0,0)$ and $\boldsymbol{\Omega}_0=(0,0,\Omega)$ are the MRF linear and angular velocities in the stationary frame, and $\mathbf{f}$ and $\rho$ are the forcing term and density respectively. Equation~\ref{eq:MRF} is discretised in the x-y plane using spectral/hp discretisation (Fig~\ref{fig:schematic}(a)). In the z-dimension, it is computationally efficient for DNS to only model a section of the VAWT structure via a homogeneous extrusion with spanwise periodic boundaries. This can be efficiently represented within the Nektar\texttt{++} framework by a Fourier expansion (Fig~\ref{fig:schematic}(b)), and here the extrusion length is set to $L_z/c=0.2$. The solution is obtained over the computational domain as illustrated in Fig.~\ref{fig:schematic}, defined by a circular far-field with a high-order convective outflow enforced through a Robin boundary condition A high-order velocity correction scheme (VCS) \citep{karniadakisSpectralHpElement2005} is then applied to Eq.~\ref{eq:MRF} to temporally advance the solution. The non-dimensional timestep was $\Delta t^*= \Delta t \omega \sim \mathcal{O}(10^{-4})$ in order to satisfy $CFL < 1$. The mesh for the domain was generated using the Nektar\texttt{++}  high-order mesh generation package \textit{NekMesh} \citep{greenNekMeshOpensourceHighorder2024}. The meshing process was controlled to ensure the initial non-dimensional grid spacings at the turbine geometry were kept to the recommended values of $\Delta x^+\leq20$, $\Delta y^+\lesssim1 $, and $\Delta z^+\leq10 $ \citep{georgiadisLargeEddySimulationCurrent2010}. Achieving DNS resolution required 32 Fourier planes in the spanwise direction and a high-order boundary layer mesh with $\Delta y/c=0.005$, which was generated using \textit{NekMesh}  \citep{greenNekMeshOpensourceHighorder2024}. Whilst DNS is considered everywhere, some regions may become slightly under-resolved, in which case Spectral Vanishing Viscosity (SVV) is employed for numerical stability, acting to damp the highest-frequency modes. Simulations are first run using polynomial order $\mathcal{P}=2$ for ten rotations and the resulting flow field is then used as the initial condition for subsequent $\mathcal{P}=4$ runs. Typically, the first couple of rotations with high-order $\mathcal{P}$ still displayed some initial effects, and so these were also discarded. The subsequent ten rotations with $\mathcal{P}=4$ are used here. The data was then averaged across rotations to gain a phase-averaged rotation. Additionally, data within Section \ref{sec:results} is also averaged across the spanwise dimension.
	
	\subsection{Validation}
	\label{sec: validation}
	
	\begin{figure}[!hbtp]
		\centerline{\includegraphics[width=0.9\textwidth]{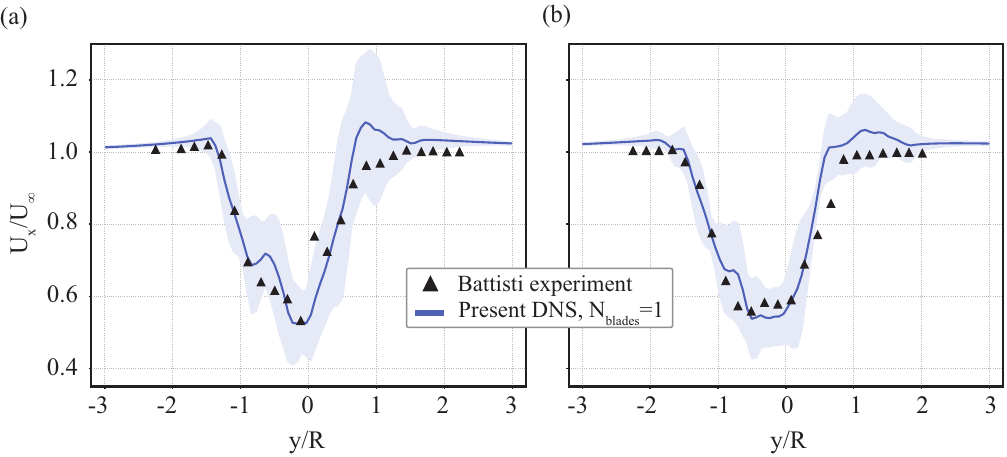}}
		\caption{Cross-stream profiles of the time-averaged streamwise velocity at (\textit{a}) $x/R = 0.5$ and (\textit{b}) $x/R = 2.0$ for $\lambda = 1.5$. Present DNS with $N_{blades}=1$ (blue line), experimental data of \cite{battistiExperimentalBenchmarkData2018} (triangles).}
		\label{fig:validation_solidity}
	\end{figure}
	
	The sparsity of available data at $Re \sim \mathcal{O}(10^4)$ made validation challenging. Much of the available data at $Re=10^4$ is heavily influenced by blockage effects. For instance, \cite{brochierWaterChannelExperiments1986} saw blockage of $\epsilon=\frac{A_s}{WH}=60\%$, where $A_s$, $W$, $H$ are the VAWT swept area, and the wind tunnel width and height respectively. This precluded it from present use, where blockage effects are minimal with $\epsilon < 1\%$. Instead, the experimental results by \cite{battistiExperimentalBenchmarkData2018} have been utilised, where their three-bladed VAWT led to $\epsilon\,<\,10\%$. The one-bladed VAWT described in Section \ref{sec: problem description} is plotted here because its solidity of $\sigma = 0.2$ is the most similar to the set-up in \cite{battistiExperimentalBenchmarkData2018}, which had $\sigma=0.25$. Another difference is the experiments were conducted under $Re\sim\mathcal{O}(10^5)$, an order of magnitude greater than the numerical data. Despite these differences, Fig.~\ref{fig:validation_solidity} shows strong alignment between the experimental and numerical data for the normalised streamwise velocity profiles, $U_x/U_\infty$. These are presented at two downstream locations, $x/R=0.5, 2.0$. At $x/R=0.5$, the strongest agreement is seen on the windward side ($y/R<0$), where the sharp gradients associated with the shear-layer at the wake edge overlap, as does the subsequent plateau for $y/R < -1.5$. The minimum values also coincide, and portray windward asymmetry. On the leeward side, the numerical data experiences an acceleration not seen experimentally. Section \ref{subsec:blade_number} discusses that this is a by-product of the DSV, which for the one-bladed case passes into the leeward side of the wake as a large coherent structure, causing flow acceleration (Fig.~\ref{fig:contour_b1}). Conversely, as solidity increases, the DSV breaks down prior to the near wake (Fig.~\ref{fig:contour_b3}). This earlier breakdown is consistent with the absence of the flow acceleration in the experimental data. Downstream for $x/R=2.0$, the plateau locations largely overlap on the windward side. Further, the downwind transition from the asymmetric velocity deficit towards a more Gaussian profile is captured numerically, with the minimum values again overlapping. Similar accuracy has been reported within explicit LES literature \cite{sheidaniAssessmentURANSMethods2023}. The strong agreement at two downstream locations indicate methodology's suitability for predicting VAWT flows.
	
	\section{Results}
	\label{sec:results}

	\subsection{Dynamic Stall and Blade-Vortex Interactions}
	\label{subsec:blade_number}
	
	\begin{figure}[!hbtp]
		\centering
		\includegraphics[width=\textwidth]{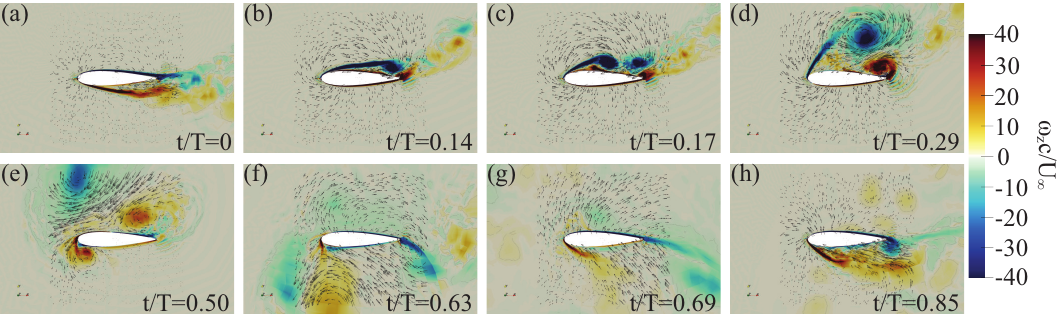}
		\caption{Phase-averaged normalised spanwise vorticity $\omega_z^*=\omega_z c/U_\infty$ close to the blade for the one-bladed configuration at $\lambda=1.5$. Subplots (a)--(h) correspond to phases $t/T=0, 0.14, 0.17, 0.29, 0.50, 0.63, 0.69, 0.85$ respectively. The black arrows depict the velocity vector.}
		\label{fig:blade_closeup}
	\end{figure}
	
	Figure \ref{fig:blade_closeup} shows the dynamic stall process occurring over the aerofoil for the one-bladed configuration at $\lambda=1.5$. This process occurs twice per rotation. First, during the upwind period, it takes place over the aerofoil surface which is closest to the shaft, named here the inner surface (Fig.~\ref{fig:blade_closeup}(a--d)). As the blade approaches the downwind region, the suction side and the separation process shifts towards the outer surface (Fig.~\ref{fig:blade_closeup}(e--h)). At $t/T=0$, the inner surface sees a boundary layer across its chord, which begins to roll-up ($t/T=0.14$), forming a coherent DSV structure, with vorticity accumulation towards the leading edge ($t/T=0.17$). During this period a new boundary layer develops downwind of the DSV, which subsequently rolls up into a separate shear layer vortex (SLV) towards the trailing edge. Subsequently, the SLV is convected away, and the DSV spreads across the chord, leading to a massively separated flow surrounding the inner surface ($t/T=0.29$). Further, the region underneath the DSV is prone to secondary separation, as indicated by the positive vorticity in Fig.~\ref{fig:blade_closeup}(d) . This is consistent with the results of \citep{visbalAnalysisDynamicStall2018}. It is also similar to what was seen experimentally by \cite{lefouestDynamicStallDilemma2022}, who presented a one-bladed configuration at $\lambda=1.5$, with the solidity matching the present case. However, one noticeable difference to the experimental results for the upwind passage is the separation over the outer surface at $t/T=0$, while \cite{lefouestDynamicStallDilemma2022} notes an attached flow at the onset of the upwind zone. Examination of Fig.~\ref{fig:blade_closeup}(e--h) shows this separation is the remnants of the dynamic stall process in the downwind passage. As the experimental $Re$ was five-times greater, this may have led to more rapid downstream convection of this positive vorticity, allowing for subsequent reattachment. Regardless, as the dynamic stall process throughout Fig.~\ref{fig:blade_closeup}(a--d) shows good agreement, this outer surface behaviour is deemed to have negligible effects on the inner surface dynamic stall. 
	
	\begin{figure}[!hbtp]
		\centering
		\includegraphics[width=\textwidth]{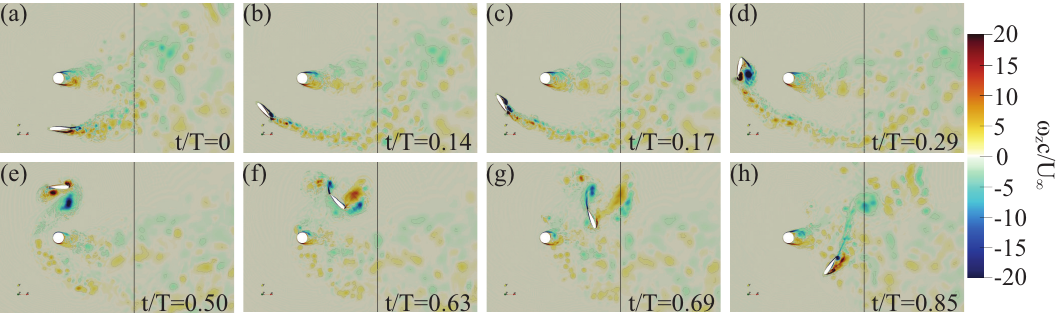}
		\caption{Normalised phase-averaged spanwise vorticity $\omega_z^*=\omega_z c/U_\infty$ over the rotor and near-wake region for the one-bladed configuration at $\lambda=1.5$ for (a)--(h) corresponding to phases $t/T=0, 0.14, 0.17, 0.29, 0.50, 0.63, 0.69, 0.85$ respectively. Vertical black line corresponds to $x/R=0.5$, as seen in the validation.}
		\label{fig:contour_b1}
	\end{figure}
	
\begin{figure}[!hbtp]
		\centering
		\includegraphics[width=\textwidth]{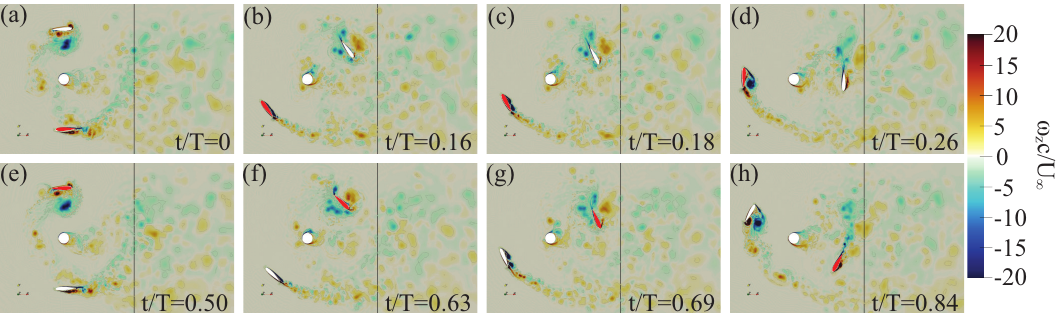}
		\caption{Normalised phase-averaged spanwise vorticity $\omega_z^*=\omega_z c/U_\infty$ over the rotor and near-wake region for the two-bladed configuration at $\lambda=1.5$ for (a)--(h) corresponding to phases $t/T=0, 0.16, 0.18, 0.26, 0.50, 0.63, 0.69, 0.84$ respectively. Vertical black line corresponds to $x/R=0.5$, as seen in the validation. Note that the text refers to the aerofoil coloured red.}
		\label{fig:contour_b2}
	\end{figure}

\begin{figure}[!hbtp]
		\centering
		\includegraphics[width=\textwidth]{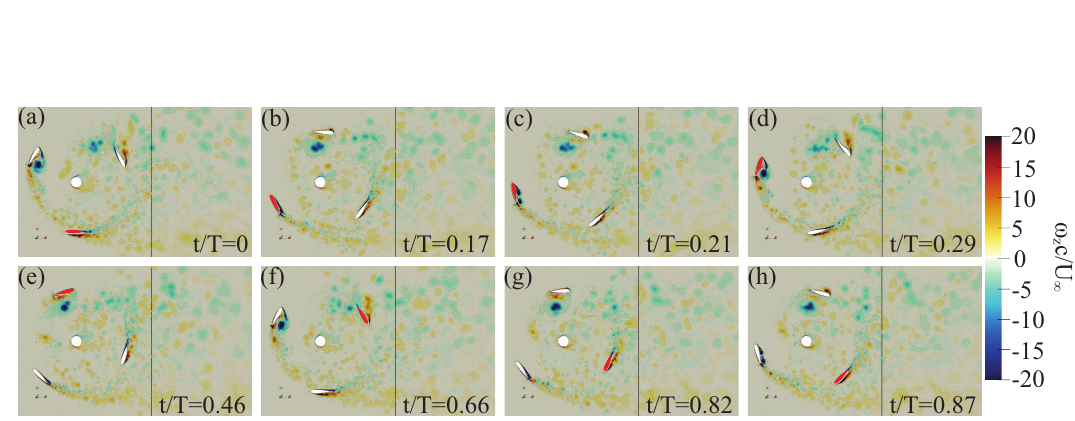}
		\caption{Normalised phase-averaged spanwise vorticity $\omega_z^*=\omega_z c/U_\infty$ over the rotor and near-wake region for the three-bladed configuration at $\lambda=1.5$ for (a)--(h) corresponding to phases $t/T=0, 0.17, 0.21, 0.29, 0.46, 0.66, 0.82, 0.87$ respectively. Vertical black line corresponds to $x/R=0.5$, as seen in the validation. Note that the text refers to the aerofoil coloured red.}
		\label{fig:contour_b3}
	\end{figure}

\begin{figure}[!hbtp]
		\centering
		\includegraphics[width=\textwidth]{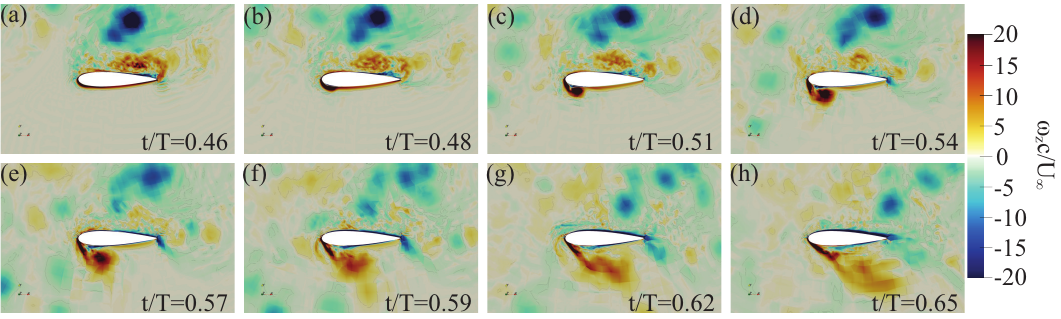}
		\caption{Phase-averaged normalised spanwise vorticity $\omega_z^*=\omega_z c/U_\infty$ close to the blade for the three-bladed configuration at $\lambda=1.5$. Subplots (a)--(h) correspond to phases $t/T=0.46, 0.48, 0.51, 0.54, 0.57, 0.59, 0.62, 0.65$ respectively.}
		\label{fig:closeup_b3}
	\end{figure}

\begin{figure}[!hbtp]
		\centering
		\includegraphics[width=\textwidth]{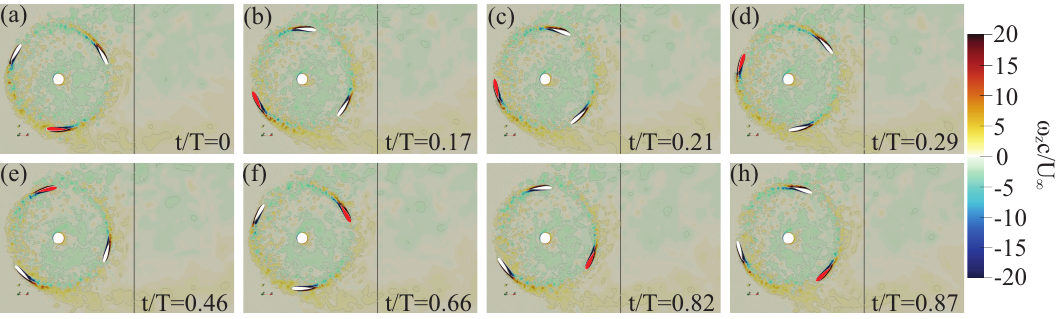}
		\caption{Normalised phase-averaged spanwise vorticity $\omega_z^*=\omega_z c/U_\infty$ over the rotor and near-wake region for the three-bladed configuration at $\lambda=4.5$ for (a)--(h) corresponding to phases $t/T=0, 0.17, 0.21, 0.29, 0.46, 0.66, 0.82, 0.87$ respectively. Vertical black line corresponds to $x/R=0.5$, as seen in the validation. Note that the text refers to the aerofoil coloured red.}
		\label{fig:contour_tsr4p5}
	\end{figure}

	To view how this process manifests across the rotor region, Fig.~\ref{fig:contour_b1}(a--d) shows the upwind dynamic stall occurring from the VAWT's perspective for the one-bladed configuration. After the DSV detaches from the blade ($t/T=0.50$), it is convected in front of the leading edge, and enters the near-wake as a coherent structure. Further, the outer surface leading edge separation also is convected into the near-wake as a single structure. Finally, during $0.63<t/T<0.86$, a structure develops from the shed wake. Ultimately, three large-scale coherent structures enter the near-wake on the leeward side. This aligns with the validation (See Fig.~\ref{sec: validation}), where the leeward wake edge saw flow acceleration. Comparatively, the wake edge on the windward side is only influenced by the outer surface leading-edge separation ($t/T \geq 0.86$) and the shed wake, no longer being influenced by the upwind DSV. Further, for $t/T \geq 0.86$, the leading edge separation increasingly enters the near-wake as smaller structures. Hence, the windward wake edge does not witness this acceleration. The different recovery mechanism across the VAWT become apparent. The leeward side utilises coherent large-scale structures to promote cross-stream mixing \cite{stromNearwakeDynamicsVerticalaxis2022}. In contrast, the windward side is associated with a more intense shear layer \citep{stromNearwakeDynamicsVerticalaxis2022}, which has been noted to see earlier destabilisation and more rapid subsequent recovery \citep{arayaTransitionBluffbodyDynamics2017, posaDependenceWakeRecovery2020}. For the one-bladed configuration, this shear-layer is not yet particularly visible, except during the windward passage of Fig.~\ref{fig:contour_b1}(e--h); however, it does become increasingly intense as $\sigma_D$ increases, as in Figs.~\ref{fig:contour_b2}-\ref{fig:contour_tsr4p5}. The connection between the formation, separation, and convection of the DSV with the near-wake dynamics is consistent with current numerical LES \citep{posaLargeEddySimulation2018, posaDependenceWakeRecovery2020} and experimental literature \citep{arayaTransitionBluffbodyDynamics2017}.
	
	Increasing to $N_\text{blades}=2$, the upwind dynamic stall process remains largely unchanged (Fig.~\ref{fig:contour_b2}(a--d)). This is attributed to the clean fluid the leading edge still experiences; at this solidity the blade spacing remains sufficient that BVI does not significantly influence the upwind dynamic stall process. However, the behaviour post-separation is different. Firstly, Fig.~\ref{fig:contour_b2}(e--f) shows the DSV is no longer transported in front of the leading edge, remaining in the rotor region for a greater period. A likely cause for this decreased DSV convection is the twice-as-large static solidity. This higher solidity results in less freestream fluid entering the rotor, reducing vortex convection. Therefore, prior to the near-wake, the DSV  undergoes increased diffusion, breaking down into smaller-scale structures, with multiple vortex cores visible in Fig.~\ref{fig:contour_b2}(f).

	The increased solidity also reduces the advection of the outer-surface leading-edge separation into the near-wake, as well as the extent of the shed wake. In particular, the rate of advection of the outer-surface separation is noticeably decreased with increasing blade number. For the one-bladed configuration, this passes into the VAWT near-wake by $t/T=0.69$, whilst this has not occurred for the two-bladed VAWT (Fig.~\ref{fig:contour_b1}(g) and (Fig.~\ref{fig:contour_b2}(g) respectively). Furthermore, the shed wake that develops during the downwind passage for the one-bladed configuration (Fig.~\ref{fig:contour_b1}(f--h)) exhibits a greater spatial extent, leading to a more pronounced influence on the near-wake. In contrast, the two-bladed configuration produces a comparatively weaker and less extensive shed wake (Fig.~\ref{fig:contour_b2}(f--h)). Collectively, these behaviours contribute to the reduced flow acceleration observed with increasing $N_\text{blades}$ at the leeward near-wake edge, as discussed for Fig.~\ref{fig:validation_solidity} in Section~\ref{sec: validation}.
	
	Increasing to $N_\text{blades}=3$, the reduced blade spacing introduces BVI that were absent for $N_\text{blades}=2$. The three-bladed configuration experiences significant BVI at the onset of the upwind passage due to this reduced spacing (Fig.~\ref{fig:contour_b3}(a)). The blade then experiences a relatively clean fluid for $0.17 \leq t/T \leq 0.46$ (Fig.~\ref{fig:contour_b3}(b-e). However, these initial interactions occur during the early stages of the dynamic stall process, which is characterised by laminar separation bubble formation, growth, and eventual breakdown prior to large-scale separation \citep{visbalAnalysisDynamicStall2018}. Previous studies have demonstrated that BVI can directly interact with LSB behaviour, promoting instability and modifying transition into leading-edge separation \citep{barnesCounterclockwiseVorticalGustAirfoil2018}. Accordingly, the presence of BVI at this early phase likely perturbs the LSB development and breakdown process, and additionally modifies the initial conditions from which the DSV emerges. This provides a mechanism for the reduced DSV size observed for $N_\text{blades}=3$ (Fig.~\ref{fig:contour_b3}(e)), highlighting how BVI introduced at the onset of dynamic stall can influence the subsequent evolution of the entire stall cycle. To the best of the author's knowledge, such a mechanism has not previously been extended to VAWT configurations.
	
	Beyond the upwind passage, the three-bladed rotor also experiences BVI as the blade approaches the downwind zone ($t/T \lesssim 0.5$) (Fig.~\ref{fig:contour_b3}(e)). In contrast, for $N_{\text{blades}}=1$ and $2$ the interactions were confined to $t/T > 0.5$, after the blade had entered the downwind region. The additional BVI arise from the reduced blade spacing and the increasingly disturbed flow field, both of which intensify with $N_\text{blades}$. These are highlighted in Fig.~\ref{fig:closeup_b3}, which shows the period within Fig.~\ref{fig:contour_b3}(e--f), in increments of $\Delta t \approx 0.03T$. Here, the upcoming wake is now populated with vortical structures Fig.~\ref{fig:closeup_b3}(a, b). These upstream vortices collide with the DSV Fig.~\ref{fig:closeup_b3}(c--e), causing it to break up into multiple distinct cores Fig.~\ref{fig:closeup_b3}(f--h). The interactions also decelerate the DSV relative to the blade, producing a compounding effect, where the vortex remains within the rotor region for longer, and therefore experiences further interactions. Whilst vortex interaction effects on the DSV have been discussed for isolated aerofoils \citep{barnesCounterclockwiseVorticalGustAirfoil2018}, and are known to be significant for VAWT \cite{posaLargeEddySimulation2018}, to the best of the author's knowledge, this is the first time vortex impingement has been shown to cause the premature breakup of the DSV within VAWT. This aids in understanding why increasing blade number leads to smaller vortical structures in the near-wake as seen in \cite{arayaTransitionBluffbodyDynamics2017}.
	
	At $\lambda=4.5$, an aerofoil for a three-bladed rotor no longer exhibits dynamic stall at any point of its rotation (Fig.~\ref{fig:contour_tsr4p5}). The higher TSR reduces the AoA range (see Section \ref{sec: problem description}), preventing the overshoot past the static stall angle. Hence, separation is only experienced at the trailing edge, with no vorticity accumulation at the leading edge, or coherent DSV. Instead, the flow over the aerofoil shows much less variation with rotation angle. The most noticeable variation is the more intense shed wake, which contains additional vortical, during the windward passage (Fig.~\ref{fig:contour_tsr4p5}(a--b)) compared to the leeward (Fig.~\ref{fig:contour_tsr4p5}(e--f)). This is likely attributable to the higher relative velocity over the blade during the windward period. Further, the turbine begins to act as a bluff body, with reduced fluid travelling through the VAWT structure into the near-wake. This can be seen from the vortical structures occupying the full width of the near-wake for the three-bladed VAWT with $\lambda =1.5$ (Fig.~\ref{fig:contour_b3}). In contrast, as TSR increases to $\lambda=4.5$ (Fig.~\ref{fig:contour_tsr4p5}), these are largely confined to the shear layers at the wake edges. This transition to bluff-body dynamics is consistent with \cite{arayaTransitionBluffbodyDynamics2017}. Accordingly, the near-wake experiences diminished influence from coherent vortical structures such as the DSV. Instead, it becomes increasingly defined by the wake shear-layers, with recovery is driven by shear-layer-associated mechanisms, consistent with \citep{arayaTransitionBluffbodyDynamics2017, posaDependenceWakeRecovery2020}.

	Taken together, the $\lambda=4.5$ case and the $N_\text{blades}=3$, $\lambda=1.5$ case approach bluff-body dynamics from two distinct directions. At $\lambda=4.5$, the transition follows the route identified by \cite{arayaTransitionBluffbodyDynamics2017}. Here, the reduced AoA range suppresses dynamic stall entirely, and the rotor's increased dynamic solidity limits freestream penetration, leading to the wake shear-layers driving recovery. The progression from one-bladed to three-bladed rotors at $\lambda=1.5$ reveals a complementary mechanism. The BVI-driven breakdown of the DSV removes the coherent structures that would otherwise distinguish the VAWT wake from that of a bluff body, accelerating the transition independently of the blockage effect. The downstream consequences of this are examined in the next section.
	
	\subsection{Wake Characteristics and Self-Similarity}
	\label{subsec:self_similarity}
	
	\begin{figure}[!hbtp]
		\centering
		\includegraphics[width=\textwidth]{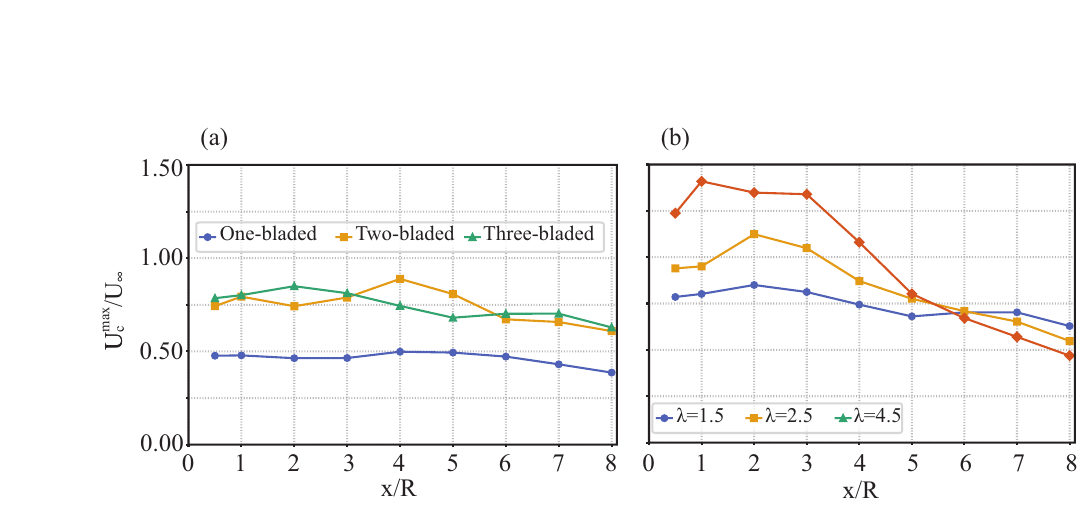}
		\caption{Downstream evolution of the normalised maximum velocity deficit $U_{c}^\text{max}/U_\infty$, for (a) $N=1,2,3$ with $\lambda=1.5$ and (b) $\lambda=1.5, 2.5, 4.5$ for $N=3$.}
		\label{fig:deficit}
	\end{figure}
	
\begin{figure}[!hbtp]
		\centering
		\includegraphics[width=\textwidth]{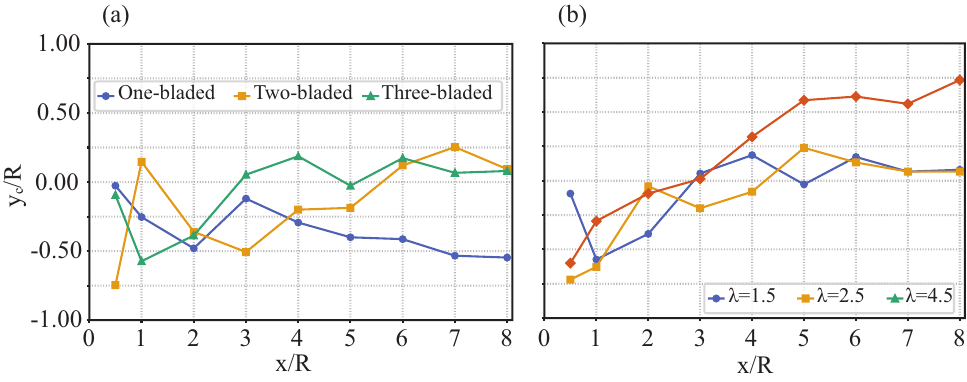}
		\caption{Downstream evolution of the normalised wake centre $y_c/R$. Shown for (a) $N=1,2,3$ with $\lambda=1.5$ and (b) $\lambda=1.5, 2.5, 4.5$ for $N=3$.}
		\label{fig:wake_centre}
	\end{figure}

	To characterise the downstream evolution of the wake, time-averaged streamwise velocity profiles are extracted up to $x/R \leq 8.0$ downstream of the rotor centre. From each profile, two quantities are extracted and presented in Fig.~\ref{fig:deficit} and Fig.~\ref{fig:wake_centre}. The first of these quantities is the normalised maximum velocity deficit $U_{c}^\text{max}/U_\infty = \max(U_c)/U_\infty = \max(U_\infty - u_x)/U_\infty$, which is therefore $1$ when the wake is completely de-energised, and $0$ when it has fully recovered. The second quantity is the wake centre $y_c$, which is defined as the cross-stream location at which $U_{c}^\text{max}/U_\infty$ occurs \cite{abkarSelfsimilarityFlowCharacteristics2017}. As expected, Fig.~\ref{fig:deficit}(a) shows more severe velocity deficits are experienced at higher $\sigma$. In line with Eq.~\ref{eq:dynamic-solidity}, this is because the decreased distance between blades results in less freestream fluid penetrating through the turbine structure and passing into the near-wake. However, high $\sigma$ also sees the recovery process initiate at an earlier downstream location, beginning at $x/R\approx2.0$ for $N_\text{blades}=3$ compared to $x/R\approx4.0, 5.0$ for $N_\text{blades}=2$ and $1$ respectively. Additionally, the recovery rate for $N_\text{blades}=1$ is much reduced compared to $N_\text{blades}=2$ and $3$, where the rate is more similar. This trend continues in Fig.~\ref{fig:deficit}(b), where increasing $\lambda$ (and therefore $\sigma_D$) further increases the initial velocity deficits, even producing recirculation in the most extreme $\lambda=4.5$ case. Recirculation regions have been previously reported with VAWT wakes at higher $\sigma_D$ \citep{arayaTransitionBluffbodyDynamics2017}. Recovery rate also becomes more rapid and initiates further upstream with increasing $\lambda$, beginning as early as $x/R\approx1$ for the most extreme case. The behaviour within Fig.~\ref{fig:deficit} suggests the transition to bluff-body dynamics, which occurs earlier at higher $\sigma_D$, is the more influential recovery mechanism. The shear-layer mechanism associated with this transition drives more rapid recovery \citep{arayaTransitionBluffbodyDynamics2017, posaDependenceWakeRecovery2020} than the cross-stream mixing induced by DSVs \citep{stromNearwakeDynamicsVerticalaxis2022} that governs the low-$\sigma_D$ regime. These two regimes can be qualitatively contrasted in the contour visualisations, with the DSV-dominated near-wake at low-$\sigma_D$ in Fig.~\ref{fig:contour_b1}, and the shear-layer-dominated wakes at high-$\sigma_D$ in Figs.~\ref{fig:contour_b3}--\ref{fig:contour_tsr4p5}.
	
	Figure~\ref{fig:wake_centre}(a) indicates extreme variability in wake centre across $N_{blades}$ up to $x/R\approx3$. However, beyond this region, a divergence in behaviour emerges between blade-number configurations. The higher solidity configurations ($N_{\text{blades}}=2,3$) see the wake centre transported towards $y_c/R=0$, whereas the one-bladed case continues to meander asymmetrically in the negative $y$-direction. From inspection of the vorticity contours in Figs.~\ref{fig:contour_b1}-\ref{fig:contour_b3}, this may be attributed to the persistence of the windward shear layer. For $N_{\text{blades}}=1$, this shear layer is only visible during the windward half of the rotation cycle and diffuses during the leeward passage, whereas for $N_{\text{blades}}=2$ and $3$ the shear layer remains increasingly visible regardless of azimuthal position. Hence, the sustained shear layer on the windward side at higher solidity configurations may act to inhibit transport of the wake centre in the windward direction, promoting its return towards the centreline. This pattern is reinforced in Fig.~\ref{fig:wake_centre}(b), where increasing $\lambda$ (and therefore $\sigma_D$) causes a progressively more rapid transport of the wake centre in the positive $y$-direction. Finally, all configurations experiencing positive $y$-transport tend towards $y_c/R=0$, with the exception of the most extreme $\lambda=4.5$ case, which substantially overshoots the centreline and ultimately exhibits greater asymmetry on the positive side than was ever observed on the negative side.
	
\begin{figure}[!hbtp]
		\centering
		\includegraphics[width=\textwidth]{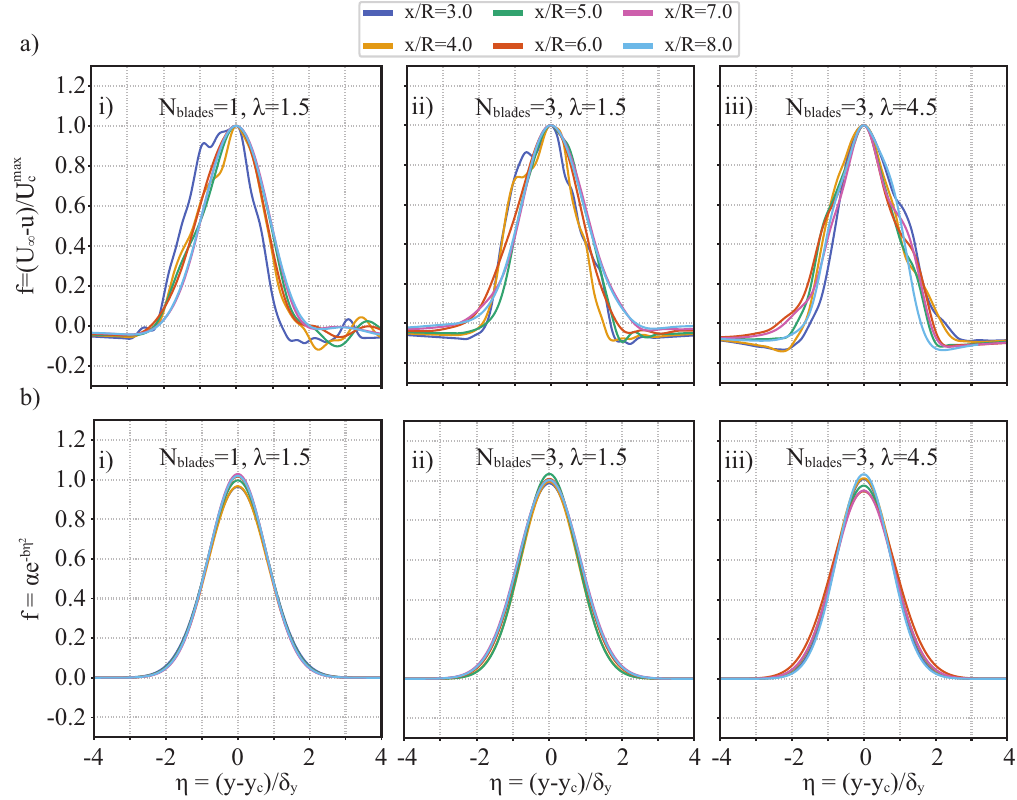}
		\caption{Scaled velocity deficit profiles $f = U_c / U_c^\text{max}$ with similarity variable $\eta = (y - y_c)/\delta_y$ at six downstream locations (a). Shown for for the one-, and three-bladed configuration with $\lambda=1.5$ (i,ii) and the three-bladed configuration with $\lambda=4.5$ (iii). Additionally plotted are the corresponding fitted functional forms (b). Subplots (i--iii) follow the same layout as (a).}
		\label{fig:selfsim}
	\end{figure}

\begin{figure}[!hbtp]
		\centering
		\includegraphics[width=\textwidth]{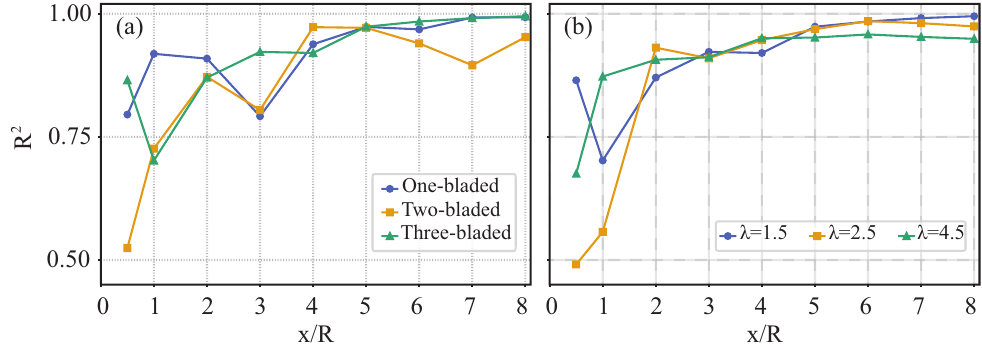}
		\caption{Coefficient of determination $R^2$ between the scaled velocity deficit profiles and the fitted functional form, as a function of downstream distance $x/R$. Shown for (a) $N_{blades}=1,2,3$ with $\lambda=1.5$ and (b) $\lambda=1.5, 2.5, 4.5$ for $N_{blades}=3$.}
		\label{fig:r2}
	\end{figure}

	The preceding trends indicate that the persistence of the DSV is closely tied to the slower recovery observed at low $\sigma_D$, whereas configurations in which its influence is diminished exhibit earlier and more rapid recovery. By conducting a self-similar analysis across the parameter space, the downstream distance at which the scaled deficit profiles collapse onto a common functional form provides a measure of how rapidly the wake loses its dependence on blade-generated coherent structures. Following \cite{abkarSelfsimilarityFlowCharacteristics2017}, the velocity deficit profiles are scaled by defining the similarity variable
	$\eta = (y - y_c)/\delta_y$, where $\delta_y$ is the wake half-width, defined as the cross-stream distance from $y_c$ at which the deficit falls to half of its maximum value. The normalised deficit is
	then given by $f = U_c / U_c^\text{max}$. These scaled profiles are presented for the $N_\text{blades}=1$ configuration at $\lambda=1.5$, and $N_\text{blades}=3$ at $\lambda=1.5, 4.5$ in Figs.~\ref{fig:selfsim}(a, i--iii) respectively. Further, the corresponding functional form $f = \alpha e^{-b\eta^2}$ \citep{tennekesFirstCourseTurbulence1972}, where $\alpha$ and $b$ are fitted iteratively, is shown in Figs.~\ref{fig:selfsim}(b, i--iii) respectively. Similarly to the findings of \cite{abkarSelfsimilarityFlowCharacteristics2017}, a persistent limitation to self-similarity is observed on the leeward wake edge, where the scaled profiles are inconsistent with the functional form. This leeward deviation decreases with increasing dynamic solidity, suggesting it is connected to the passage of the DSV. Further, closer to the rotor ($x/R = 3.0, 4.0$), notable variation is observed near the peak of the deficit profile. Since \cite{abkarSelfsimilarityFlowCharacteristics2017} only considered $x/R \geq 7$, this deviation was not captured in their analysis. Similarly, this feature also diminishes with increasing dynamic solidity. Therefore, both features limiting self-similarity are diminished at higher $\sigma_D$, indicated by a more rapid collapse onto the functional form. This is quantified in Fig.~\ref{fig:r2}(a) using the coefficient of determination $R^2 = 1 - SS_\text{res}/SS_\text{tot}$, where $SS_\text{res} = \sum(f_i - \hat{f}_i)^2$ is the sum of squared residuals between the scaled profile data and the fitted functional form, and $SS_\text{tot} = \sum(f_i - \bar{f})^2$ is the total sum of squares about the mean of the data. Here, the three-bladed configuration sees a consistent increase in $R^2$ with downstream distance for $x/R \geq 1$(Fig.~\ref{fig:r2}(a)). In contrast, $N_{\text{blades}}=1$ and $2$ experiences downstream regions where the correlation becomes weaker. This trend continues with higher TSR in Fig.~\ref{fig:r2}(b), with $\lambda=2.5$ and $\lambda=4.5$. Additionally, the strongest gradient is observed at $\lambda = 2.5$, suggesting the rate of increase in $R^2$ may diminish beyond a specific dynamic solidity. Whilst self-similarity was demonstrated at two tip-speed ratios by \cite{abkarSelfsimilarityFlowCharacteristics2017}, this was only shown for $x/R \geq 7$, likely due to the ALM employed. By incorporating the near-wake data, the present results capture the transition to the self-similar solution, where it is found that varying blade number holds significant influence on the rate of transition.
	
	\section{Conclusions}
	\label{sec:conclusions}
	DNS of one-, two- and three-bladed VAWT have been conducted using the high-order spectral/hp element method framework Nektar\texttt{++}. The geometrically-resolved methodology captures the full dynamic stall process, including the formation of the DSV, its separation from the blade, and convection into the near-wake. Increasing blade number introduces progressively more significant BVI which interfere with the dynamic stall process. For $N_\text{blades}=3$, these interactions are present at the onset of the upwind passage, coinciding with the early phase of dynamic stall during which the laminar separation bubble develops, which has previously shown to be sensitive to upstream vortical disturbances for isolated aerofoils \citep{barnesCounterclockwiseVorticalGustAirfoil2018}. The reduced DSV size observed for $N_\text{blades}=3$ is consistent with this sensitivity. Further, as the blade approaches the downwind zone, direct vortex impingement causes the premature breakup of the DSV into multiple cores. This is the first time, to the best of the authors' knowledge, that BVI-induced DSV breakup has been visualised for VAWT.
	
	In the near-wake, higher dynamic solidity produces more severe initial velocity deficits but earlier and more rapid recovery. Two complementary mechanisms drive the earlier transition to bluff-body dynamics at higher $N_\text{blades}$. First, reduced freestream penetration through the rotor, as captured by the timescales introduced in \cite{arayaTransitionBluffbodyDynamics2017}. Second is the BVI-driven breakdown of the structures associated with dynamic stall which would otherwise distinguish the VAWT wake from a bluff-body wake. The latter mechanism required a DNS approach as outlined here to be identified. The shear-layer-associated recovery that dominates at high $\sigma_D$ is here observed to be more rapid than the DSV-driven mechanism at low $\sigma_D$.
	
	Self-similarity analysis, extending \cite{abkarSelfsimilarityFlowCharacteristics2017} into the near-wake, captures the transition to the self-similar solution. Blade number is more influential than TSR on the rate of this transition, with the fastest rate observed at $\lambda=2.5$ rather than the maximum $\lambda$, suggesting a plateau beyond a certain $\sigma_D$. The more rapid recovery and earlier self-similarity onset at higher $\sigma_D$ have implications for closely-spaced turbine arrays, as the wake becomes amenable to analytical modelling (see \cite{abkarSelfsimilarityFlowCharacteristics2017}) at shorter downstream distances. For closey-space turbine arrays, which include the coupled-pair configurations, the present results indicate that the inflow experienced by a downstream rotor is fundamentally blade-number-dependent. Low $N_\text{blades}$ coupled-pairs will experience large coherent DSV compared to a diffuse, disturbed field at high $N_\text{blades}$. DNS of paired configurations is warranted to determine how these distinct inflow conditions affect the downstream rotor's flow behaviour and performance.
	
	\begin{acknowledgement}
		This work was supported by the Engineering and Physical Sciences Research Council [EP/W524700/1]. Additionally, the authors would like to thank EPSRC for the computational time made available on the UK supercomputing facility ARCHER/ARCHER2 via the UK Turbulence Consortium (EP/X035484/1).
	\end{acknowledgement}
	
	{\small
		\bibliographystyle{spbasic}
		\bibliography{references}
	}
	
\end{document}